\documentclass{emulateapj}

\usepackage{epsfig}
\usepackage{amsmath}
\usepackage{rotating}
\usepackage{natbib}
\usepackage{enumerate}
\usepackage{multirow}
\usepackage{array}
\usepackage{appendix}
\usepackage{comment}

\def\plotonesc#1{\centering \leavevmode
\includegraphics[clip=, width=1.70\columnwidth]{#1}}
\def\plotoneh#1{\centering \leavevmode
\includegraphics[clip=, width=.95\columnwidth]{#1}}

\newcommand{\cN}[1]{\mathcal{N}}

\def\gsim{\;\rlap{\lower 2.5pt
 \hbox{$\sim$}}\raise 1.5pt\hbox{$>$}\;}
\def\lsim{\;\rlap{\lower 2.5pt
   \hbox{$\sim$}}\raise 1.5pt\hbox{$<$}\;}

\begin{document}


\title{Jupiter will become a hot Jupiter:\\
Consequences of Post-Main-Sequence Stellar Evolution on Gas Giant Planets}

\author{
David S. Spiegel\altaffilmark{1},
Nikku Madhusudhan\altaffilmark{2}
}

\affil{$^1$Astrophysics Department, Institute for Advanced Study,
  Princeton, NJ 08540}

\affil{$^2$Department of Physics and Department of Astronomy, Yale
  University, New Haven, CT 06511}

\vspace{0.5\baselineskip}

\email{
dave@ias.edu,\\
Nikku.Madhusudhan@yale.edu
}

\begin{abstract}
  When the Sun ascends the red giant branch (RGB), its luminosity will
  increase and all the planets will receive much greater irradiation
  than they do now.  Jupiter, in particular, might end up more highly
  irradiated than the hot Neptune GJ~436b and, hence, could
  appropriately be termed a ``hot Jupiter.''  When their stars go
  through the RGB or asymptotic giant branch (AGB) stages, many of the
  currently known Jupiter-mass planets in several-AU orbits will
  receive levels of irradiation comparable to the hot Jupiters, which
  will transiently increase their atmospheric temperatures to
  $\sim$1000~K or more.  Furthermore, massive planets around
  post-main-sequence stars could accrete a non-negligible amount of
  material from the enhanced stellar winds, thereby significantly
  altering their atmospheric chemistry as well as causing a
  significant accretion luminosity during the epochs of most intense
  stellar mass loss.  Future generations of infrared observatories
  might be able to probe the thermal and chemical structure of such
  hot Jupiters' atmospheres.  Finally, we argue that, unlike their
  main-sequence analogs (whose zonal winds are thought to be organized
  in only a few broad, planetary-scale jets), red-giant hot Jupiters
  should have multiple, narrow jets of zonal winds and efficient
  day-night redistribution.
\end{abstract}

\keywords{planets and satellites: Jupiter --- Sun: evolution ---
  planetary systems --- radiative transfer --- stars: evolution ---
  stars: AGB and post-AGB}

\section{Introduction}
\label{sec:intro}
The ``hot Jupiter'' class of exoplanets was not generally anticipated
prior to the discoveries of the first planets around main sequence
stars \citep{mayor+queloz1995, marcy+butler1996}.  Their existence,
however, could have been predicted long before then, since the Sun's
luminosity will increase by a factor of several thousand as it ascends
the red giant branch (RGB), thereby turning Jupiter into a hot Jupiter
in several billion years.

\begin{figure*}[t!]
\plotonesc
{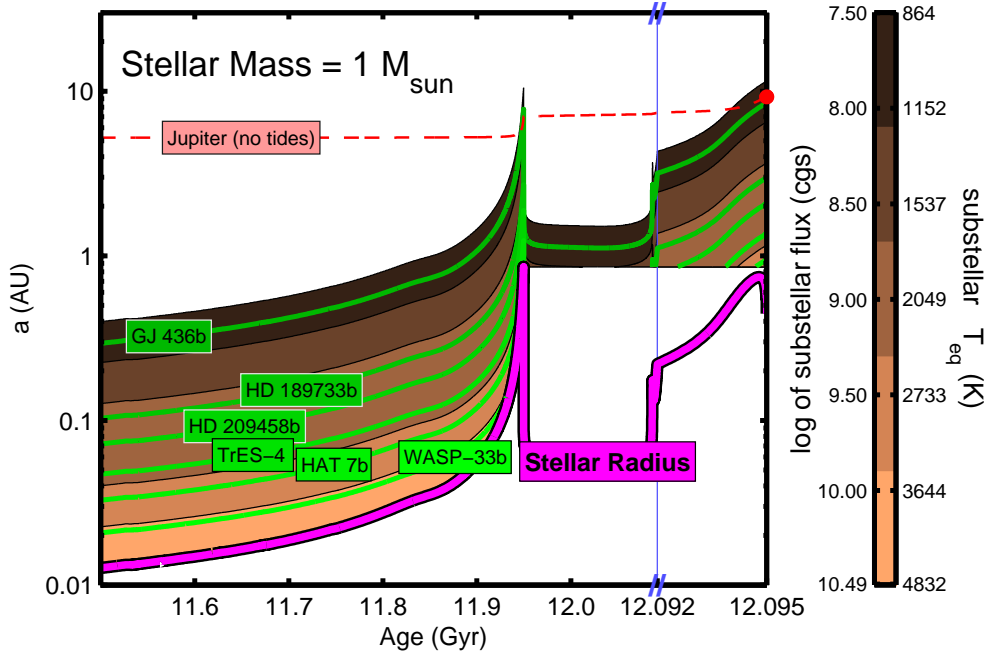}
\caption{Orbital separations where red-giant hot Jupiters can be found
  around a solar-type star.  The evolution of a 1-$M_\sun$ star's
  radius (magenta curve) is shown as a function of age in gigayears
  (Gyr), and the consequent levels of irradiation as a function of
  stellar age and orbital distance (color scale).  Green curves show
  regions of (age, orbital radius) space where an object would receive
  the same incident irradiation as known various known highly
  irradiated objects: GJ~436b, HD~189733b, HD~209458b, TrES-4,
  HAT-P-7b, and WASP-33b. The red dashed curve shows how Jupiter's
  orbit will evolve purely under the influence of stellar mass loss,
  with a red dot at the end showing the position achieved at the end
  of our stellar model.  Note that, in order to more clearly show the
  final, very brief part of the star's evolution, the time axis is
  stretched by a factor of 40 during the ascent of the AGB phase (from
  12.092~Gyr until the end of the model).  The location where the time
  axis changes scale is marked with a thin blue line and hatch marks
  along the horizontal axis.}
\label{fig:irradiation1}
\end{figure*}

Roughly 20\% of the more than 700 currently known
exoplanets\footnote{See http://exoplanet.eu
  \citep{schneider_et_al2011}, or see http://exoplanets.org
  \citep{wright_et_al2011} for a differently vetted list.} have masses
greater than half of Jupiter's, orbital radii greater than 1~AU, and
will become hot Jupiters (i.e., for the present purposes, this means
they will receive at least as much irradiation as the hot Neptune
GJ~436b) before the end of stars' lives.  Depending on the efficiency
of tidal dissipation in RGB and AGB stars, many of these planets might
eventually be tidally engulfed by their stars
\citep{carlberg_et_al2009, villaver+livio2009, nordhaus_et_al2010},
where they could play a role in shaping planetary nebulae (PNe)
\citep{soker_et_al1984, nordhaus+blackman2006, nordhaus_et_al2007} or
creating highly magnetized white dwarfs \citep{nordhaus_et_al2011}.
Regardless of whether these planets are eventually swallowed by their
stars, at some point in their futures they will be highly irradiated.

Some searches for evidence of planets and other companions to
post-main-sequence stars have already been undertaken.  White dwarf
atmospheres polluted by tidally shredded asteroids indicate the
presence of distant planetary companions around $\sim$30\% of white
dwarfs \citep{zuckerman_et_al2010}.  A post-common-envelope 50-$M_J$
(Jupiter-mass) object was found in a tight orbit around a white dwarf
\citep{maxted_et_al2006}.  Tentative evidence has also suggested the
existence of several other systems, including two small planets in
tight orbits around a subdwarf star \citep{charpinet_et_al2011}, and a
jovian body around the pulsating white dwarf GD-66
\citep{mullally_et_al2007, mullally_et_al2008, mullally_et_al2009} ---
although recent data have complicated the planetary hypothesis for the
latter system (\citealt{farihi_et_al2012}; Hermes, private
communication).  Direct-imaging searches for warm companions to white
dwarfs have yet to find any \citep{hogan_et_al2009}, but have not yet
surveyed large numbers of stars.  Finally, \citet{johnson_et_al2011}
has found a number of giant planets around slightly evolved, subgiant
stars (and some of these planets will soon become hot Jupiters).

Here, we consider the properties of ``red-giant hot Jupiters''
(RGHJs), a term that we use somewhat loosely to refer generally to hot
Jupiters around post-main-sequence stars that are on longer period
orbits than their cousins around main-sequence-star.  In
\S\ref{sec:hot}, we calculate the range of orbital distances at which
a gas giant planet around a post-main-sequence star might be
considered a hot Jupiter, and we consider possible heating of a
planet's bulk interior.  In \S\ref{sec:acc}, we examine how the
accretion of stellar wind (and perhaps rocky material) onto a RGHJ
might pollute its atmosphere.  In \S\ref{sec:chemspec}, we describe
changes in Jupiter's chemistry and spectrum that will occur as the Sun
evolves beyond the main sequence.  In \S\ref{sec:motions}, we describe
some qualitative differences between the wind patterns expected on
RGHJs and on main-sequence hot Jupiters.  In \S\ref{sec:conc}, we
conclude and speculate on the observability of RGHJs.

\section{Becoming Hot}
\label{sec:hot}
In order to evaluate where around a post-main-sequence star the
irradiation is sufficient to produce a hot Jupiter, we use stellar
evolution models with zero-age main sequence (ZAMS) masses ranging
from 1 to 3~$M_\sun$.  Our stellar models are calculated using the
``Evolve Zero-age Main Sequence (EZ) code'' \citep{paxton2004}.  For
each initial stellar mass, we calculate the star's evolution through
the main sequence and to the end of the AGB phase, using metallicity
$Z=0.02$.

Figure \ref{fig:irradiation1} shows where red-giant hot Jupiters can
be found as a function of age and orbital separation, around a
1-$M_\sun$ star.  The evolution of a solar-mass star's radius and
luminosity by showing where and when a companion would be irradiated
similarly to various known hot Jupiters.  The evolution of Jupiter's
orbit is shown, taking into account only stellar mass loss (which
makes the orbit expand as $M_*^{-1}$) and neglecting tidal
interactions between planet and star (which would generally tend to
make the orbit shrink).

\begin{figure*}[t!]
\plotonesc
{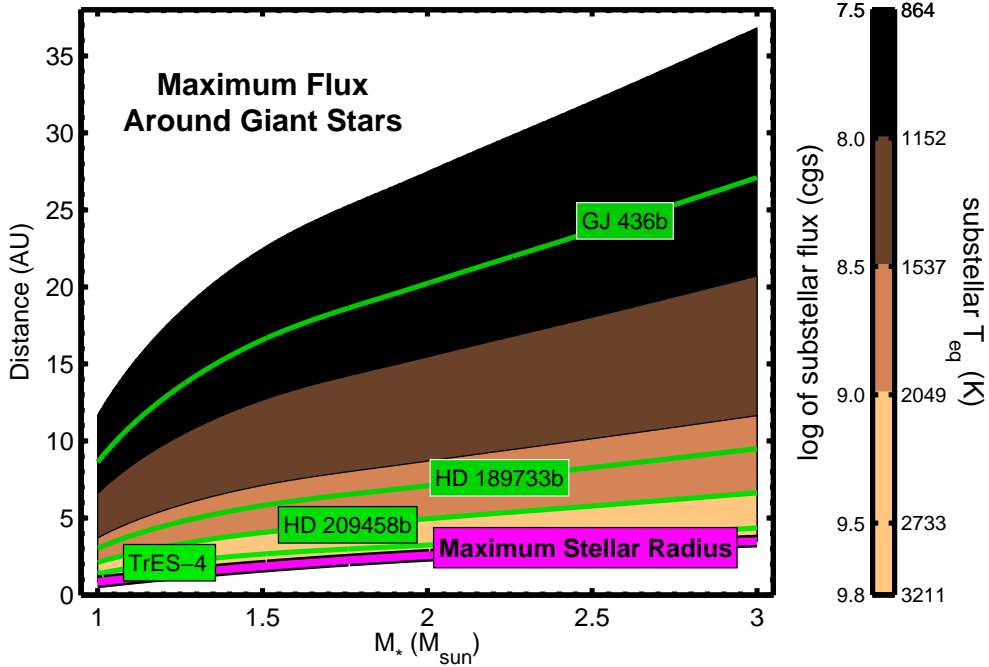}
\caption{Maximum orbital separations where red-giant hot Jupiters can
  be found around 1---3-$M_\sun$ stars.  The level of irradiation is
  shown, as a function of orbital distance and stellar mass, at the
  time of a star's maximum luminosity.  Green curves show regions of
  (stellar mass, orbital radius) space where an object would receive
  the same irradiation as several highly irradiated planets.}
\label{fig:irradiation2}
\end{figure*}

Figure \ref{fig:irradiation2} shows the maximum distance at which a
RGHJ could be found around stars ranging from 1 to 3 times the mass of
the Sun, at the time when the stars will be maximally luminous.  A
companion could (briefly) receive irradiation of $\sim$$3\times 10^{7}
\rm~erg~cm^{-2}~s^{-1}$ as far out as $\sim$12~AU around a ZAMS
1-$M_\sun$ star (during the RGB phase) or as far out as $\sim$35~AU
around a 3-$M_\sun$ star (during the AGB phase).

The detailed shape of the shaded ``hot'' contours in
Figs.~\ref{fig:irradiation1} and \ref{fig:irradiation2} depends on
properties of stars (including metallicity and mass loss) that are
surely not perfectly captured by our models, but the qualitative
features of our models should hold for real stars.  Planets at a given
hot region in either the left or the right panel of this figure could
have had main-sequence orbital separations that were either greater or
less than the given abscissa -- greater if tidal interactions are
reducing the orbital separation, less if stellar mass loss is causing
it to expand.\footnote{When the star is early on the subgiant branch,
  its angular rotation rate might not have decreased by much yet, and
  tidal interactions would actually cause companions whose orbital
  mean motion is less than the stellar rotation rate to move {\it out}
  (see, e.g., \citealt{lloyd2011}), similar to the Earth-Moon
  interaction.}  We note that more post-main-sequence hot Jupiters in
a volume-limited survey will be found at moderate separation (one to a
few AU) around subgiant and early giant branch stars than at more
extreme separation around stars at the tip of the RGB or the AGB,
simply because of the relative timescales of the respective stages of
stellar evolution.

Finally, a jovian planet that is sufficiently highly irradiated might
achieve non-negligible ionization in its atmosphere.  Winds crossing
the planetary magnetic field could, therefore, induce an electric
field that would drive currents in the atmosphere or in the deep
interior, similar to an analogous process in main sequence hot
Jupiters \citep{batygin+stevenson2010, perna_et_al2010a,
  perna_et_al2010b, batygin_et_al2011, menou2012}.  This process might
be able to contribute to heating the bulk interior of a
planet.\footnote{We note, though, that \citet{wu+lithwick2012} argue
  that Ohmic dissipation cannot reinflate objects that have already
  cooled.}  In this way, Ohmic heating (or any other mechanism that
converts irradiation energy into thermal energy in the deep interior),
if viable, could ``reset'' a planet's evolutionary clock, analogously
to the process described by \citet{ibgui+burrows2009}.  The degenerate
stellar core that is left at the end of the AGB phase is itself
extremely luminous, and might further contribute to heating (and
reheating) a planet that remains in the planetary nebula (PN), and
\citet{villaver+livio2007} consider possible evaporation of a planet
that might occur due to X-ray and ultraviolet irradiation.  A giant
planet's luminosity is generally thought to be a reasonably good
indication of its mass, irrespective of initial conditions
(``hot-start,'' ``cold-star,'' or ``warm-start''), so long as the
object is sufficiently old and its age is known
\citep{marley_et_al2007, fortney_et_al2008b, spiegel+burrows2012}.
However, if RGHJs or post-RGHJs are observed (orbiting giant stars or
the remnant white dwarfs, assuming that they did not fully evaporate
during the PN phase), it is conceivable that mass estimates based on
photometry could be biased high if the object's orbit caused it to be
sufficiently highly irradiated and if this potential reheating effect
is not taken into account.

However, there is an an important difference between RGHJs and
main-sequence hot Jupiters in terms of the timescale during which the
object is highly irradiated.  Hot Jupiters around main-sequence stars
need to convert only a small fraction of incident stellar irradiation
to a form that heats the convective interior in order to have a strong
influence on the bulk thermal state \citep{guillot+showman2002,
  liu_et_al2008}, but this is because the objects can find a
quasi-steady thermal state on a timescale of tens of millions of
years.  But the total energy intercepted by Jupiter during the Sun's
RGB and AGB phases is probably insufficient to cause significant
reheating, even if the energy were deposited deep in the convective
core, simply because Jupiter will not receive extreme irradiation for
a long enough time.  Planets that are closer to their stars receive
greater irradiation, but are at greater risk of being tidally engulfed
by their stars (potentially causing observable signatures;
\citealt{struck-marcell1988}) and therefore might not survive to the
end of the AGB phase.  Since many of the most highly irradiated RGHJs
will eventually merge with their stars, it is unclear whether any of
the objects that are on distant-enough orbits to avoid engulfment
(i.e., to end up on the outside of the circum-white-dwarf ``gap''
predicted by \citealt{nordhaus_et_al2010}) could be irradiated
sufficiently to cause any significant bulk reheating.

\section{Accreting Material}
\label{sec:acc}
As stars ascend the RGB and the AGB, they lose mass much faster than
they do when on the main sequence.  Furthermore, since wind speeds are
typically a multiple of the surface escape speed that is of order a
few times unity \citep{willson2000}, giant stars' winds are {\it much}
slower than those of main sequence stars.  For instance,
\citet{zuckerman+dyck1989} examined a sample of $\sim$100 AGB stars
and found that their outflow velocities are often less than
40~km~s$^{-1}$, and tend to cluster around 5---25~km~s$^{-1}$.  A
massive enough planet could have an escape speed that is greater than
the relative speed $v_{\rm rel}$ between the planet and the wind.  In
such a case, the planet will accrete the stellar wind material that it
passes through, with an effective ``accretion radius'' that may be
found by setting to zero the sum of the specific gravitational and
kinetic energies of wind particles, as per \citet{bondi+hoyle1944} or
\citet{dong_et_al2010}:
\begin{equation}
\label{eq:Racc} R_{\rm acc} = \frac{2 G M_p}{v_{\rm rel}^2}  = 2 a \left(\frac{M_p}{M_*} \right) \left( \frac{v_K}{v_{\rm rel}} \right)^{2} \, ,
\end{equation}
where $M_p$ and $M_*$ are the respective masses of the planet and the
star, and $a$ is the orbital semimajor axis, and $v_K$ is the
Keplerian orbital speed.  In particular, when $R_{\rm acc} \ge R_p$
(where $R_p$ is the planet's radius), the planet readily accretes
stellar wind.\footnote{Although some terrestrial planets' atmospheres
  might be eroded by main sequence stellar winds
  \citep{zendejas_et_al2010}, evolved stars' winds consist of much
  lower energy particles and will tend to accrete onto jovian-mass
  planets.} When $R_{\rm acc} < R_p$, the situation is significantly
more complicated, but even in this case some stellar wind material
will surely be intercepted and accreted.  For now, we consider planets
massive enough that $R_{\rm acc} \ge R_p$, but the qualitative
conclusions are not necessarily altered even if this condition is not
met.

If the planet accretes everything within $R_{\rm acc}$, at its
relative speed $v_{\rm rel}$, the accretion rate is
\begin{equation}
\dot{M}_p \sim \left(\pi R_{\rm acc}^2\right) \rho[a] v_{\rm rel} \, ,
\label{eq:Mp-dot}
\end{equation}
where $\rho[a]$ is the wind density at the location of the planet:
\begin{equation}
\rho[a] = -f_i \frac{\dot{M}_*}{4 \pi a^2 v_w} \, .
\label{eq:rho}
\end{equation}
In Eq.~(\ref{eq:rho}), $\dot{M}_*$ is the stellar mass-loss rate,
$v_w$ is the speed of the stellar wind, and $f_i$ is the ``wind
isotropy factor,'' 1 for a purely isotropic wind and either less than
or greater than 1 depending on if the wind is less or more dense at
the planet's position than an isotropic wind would be.  The planet's
accretion rate may, therefore, be expressed as
\begin{eqnarray}
\nonumber \dot{M}_p & \sim & -f_i \dot{M}_* \left(\frac{R_{\rm acc}}{2a}\right)^2  \left(\frac{v_{\rm rel}}{v_w}\right) \\
 & \sim & -f_i \dot{M}_* \left(\frac{M_p}{M_*} \right)^2 \left( \frac{v_K}{v_{\rm rel}} \right)^4  \left(\frac{v_{\rm rel}}{v_w}\right) \, .
\label{eq:Mp-dot2}
\end{eqnarray}
To a rough approximation, the total accreted mass is
\begin{equation}
\Delta M_p \sim |\Delta M_*| \left( \frac{M_p}{M_*} \right)^2 \left( \frac{f_i v_K^4}{v_w v_{\rm rel}^3}  \right) \, ,
\label{eq:Macc}
\end{equation}
where $\Delta M_*$ mass lost by the star.

The accreted mass can make a non-negligible contribution to the mass
of the atmosphere already present.  The mass, above pressure $P$, of
the atmosphere of a planet whose surface gravity is $g$ scales
inversely with planet mass:
\begin{equation}
\label{eq:Matm} M_{\rm atm}[P] = 4 \pi R_p^2 \frac{P}{g} = \frac{4 \pi R_p^4 P}{G M_p} \, .
\end{equation}
The ratio of the accreted mass to the mass of the atmosphere above a
given pressure level, then, is
\begin{eqnarray}
\label{eq:Matm_rat} \frac{\Delta M_p}{M_{\rm atm}[P]} & \sim &  \frac{G M_p^3 |\Delta M_*|}{4 \pi R_p^4 P M_*^2} \times \left( \frac{f_i v_K^4}{v_w v_{\rm rel}^3}  \right) \\
\nonumber & \sim & 400 \times  \left( \frac{M_p}{10 M_J}\right)^3 \left( \frac{P}{100 \rm~bars} \right)^{-1} \\
\label{eq:Matm_rat_simple} & & \times \left( \frac{|\Delta M_*|}{M_*/2} \right) \left( \frac{M_*}{M_\odot} \right)^{-1} \, .
\end{eqnarray}
In eq.~(\ref{eq:Matm_rat_simple}) we have assumed that $f_i \approx
1$, $R_p \approx R_J$, and we have taken the velocity-dependent terms
(the parenthetical term in eq.~\ref{eq:Matm_rat}) to be
$\sim$$10^{-2}$, the value appropriate for an object at 5~AU around a
solar-mass star whose wind is 40~km~s$^{-1}$.  It is worth emphasizing
that this ratio scales with the {\it cube} of the planet's mass.
Equation~(\ref{eq:Matm_rat}) shows that massive planets and brown
dwarfs could accrete a large multiple of their atmospheric masses,
causing atmospheric chemical abundances to be dominated by the
abundances of accreted stellar material.\footnote{See
  \citet{spiegel_et_al2011a} regarding the appropriateness of a
  mass-cut for distinguishing brown dwarfs from giant planets.}
Although the accreted mass might be large compared with an object's
the atmosphere mass, it is a small fraction of the object's total mass
--- of order $\sim$10$^{-5} \times (M_p/M_J)^{-2}$ --- and has a
negligible effect on the orbit \citep{villaver+livio2009}.  It is
possible that the Bondi-Hoyle accretion rate is an overestimate of the
true rate at which a companion would accrete stellar wind, since much
of the material inside $R_{\rm acc}$ but outside $R_p$, although
formally gravitationally bound to the companion, would have excess
angular momentum that must be shed before it is actually incorporated
into the companion's atmosphere.  Nevertheless, a massive enough
companion will encounter an amount of gravitationally bound stellar
wind material that is very large compared with the mass of its
atmosphere, and it is difficult to avoid the conclusion that this has
the capacity to strongly influence the companion's atmospheric
composition, especially since the atmosphere of a highly irradiated
planet develops a deep radiative layer that is statically stable and
might not efficiently mix with the deep interior of the planet
\citep{spiegel_et_al2009b}.

\begin{figure}[t!]
\plotoneh
{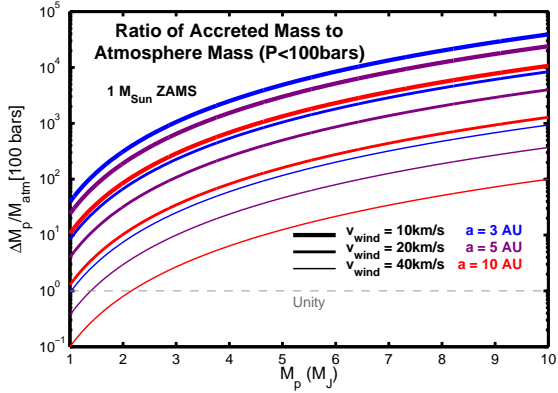}
\caption{Approxiate ratio of accreted mass to atmosphere mass above
  100~bars, for a 1~$M_\sun$ primary.  This figure is a visualization
  of equation~(\ref{eq:Matm_rat}), for a range of planet masses
  (1---10~$M_J$), orbital separations (3, 5, and 10~AU, denoted by
  colors --- blue, purple, and red, respectively), and stellar wind
  velocities (10, 20, and 40~km~s$^{-1}$, denoted by line thicknesses
  --- thick, medium, and thin, respectively).  A gray dashed line
  indicates the ordinate value 1, above which the mass of accreted
  material exceeds the mass of the atmosphere that was already
  present.}
\label{fig:accretedmass}
\end{figure}

Figure~\ref{fig:accretedmass} shows how the pollution of the planet's
atmosphere with stellar wind material scales with planet mass, stellar
wind outflow velocity, and orbital separation.  More massive planets
accrete a greater multiple of the already-present atmosphere mass
($\propto$$M_p^3$), and planet accrete more readily when closer to the
primary and when the wind velocities are lower.  For a wide range of
conditions, the planet accretes more mass than was previously present
in the atmosphere at altitudes above the 100-bar isobar.

The gravitational energy lost by stellar wind material as it falls
onto the planet constitutes a power source ($\sim$$G M_p \dot{M}_p /
R_p$).  Since $\dot{M}_p$ is proportional to $M_p^2$ (see
eq.~\ref{eq:Mp-dot2}), the accretion luminosity $L_{\rm acc}$ scales
with $M_p^3$ as follows:
\begin{eqnarray}
\label{eq:accpower1} L_{\rm acc} & \sim & -\frac{G M_p \dot{M}_*}{R_p} \left( \frac{M_p}{M_*} \right)^2 \left( \frac{f_i v_K^4}{v_w v_{\rm rel}} \right) \\
\nonumber & \sim & 10^{29} {\rm~erg~s^{-1}} \times \left( \frac{|\dot{M}_*|}{10^{-7} ~M_\sun \rm~yr^{-1}} \right) \\
\label{eq:accpower2} & & \times \left( \frac{M_p}{10 M_J} \right)^3 \left( \frac{M_*}{M_\sun} \right)^{-2} \, .
\end{eqnarray}
As in eq.~(\ref{eq:Matm_rat_simple}), we have again taken the
velocity-dependent terms to be $\sim$$10^{-2}$ (and $f_i R_J/R_p
\approx 1$).  The shocked in-falling material achieves a
characteristic temperature $T_{\rm acc}$ found by equating $k_B T_{\rm
  acc}$ to the energy lost per particle:
\begin{eqnarray}
\label{eq:falltemp1}  T_{\rm acc} & \sim & \frac{G M_p m_p}{k_B R_p} \\
\label{eq:falltemp2}  & \sim & 2\times 10^6{\rm~K} \times \left( \frac{M_p}{10 M_J} \right) \, ,
\end{eqnarray}
where $k_B$ is Boltsmann's constant and $m_p$ is the proton mass.  The
accretion process, therefore, creates an optically thin,
soft-X-ray-emitting coronal envelope around the planet, which could
produce a faint observational signature.  If the accretion power is
taken to be the X-ray luminosity, then this luminosity peaks at
$\sim$10$^{29}$~erg~s$^{-1}$ for the case of a 10-$M_J$ companion at
5~AU around a 1-$M_\sun$ (ZAMS) primary.  A more massive primary would
have a somewhat more luminous accretion shock, with luminosities
briefly reaching $\sim$10$^{30}$~erg~s$^{-1}$ for a 10-$M_J$
companion.

\begin{figure}[t!]
\plotoneh
{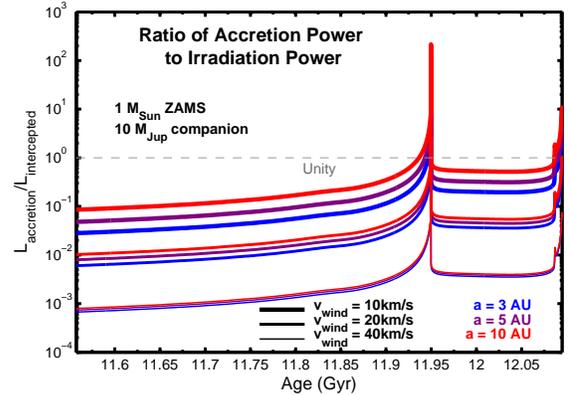}
\caption{Ratio of the accretion power to the intercepted stellar
  irradiation, for a 10-$M_J$ planet around a 1~$M_\sun$ primary.
  This figure is a visualization of equation~(\ref{eq:accpower1}), for
  a range of orbital separations (3, 5, and 10~AU, denoted by colors
  --- blue, purple, and red, respectively), and stellar wind
  velocities (10, 20, and 40~km~s$^{-1}$, denoted by line thicknesses
  --- thick, medium, and thin, respectively).  A gray dashed line
  indicates the ordinate value 1, above which the accretion power
  exceeds the irradiating power.  This occurs for only a narrow time
  window, outside of which the accretion power is only a small
  perturbation to the total heating of the planet.}
\label{fig:accretionpower}
\end{figure}

Figure~\ref{fig:accretionpower} shows the evolution of the accretion
power, calculated with eq.~(\ref{eq:accpower1}), in comparison to the
intercepted stellar irradiation.  This calculation assumes a
1-$M_\sun$ (ZAMS) primary with a 10-$M_J$ companion at a range of
orbital separations, from 3---10~AU and with a range of assumed wind
velocities, from 10---40~km~s$^{-1}$.  The accretion power exceeds the
irradiating power for only a relatively small portion of the
post-main-sequence evolution.  Since accretion power scales with the
cube of planet mass, the curves in Fig.~\ref{fig:accretionpower} would
scale up by a factor of $\sim$30 for a 30-$M_J$ companion and down by
a factor of 1000 for a 1-$M_J$ companion (such that for a Jupiter-mass
companion, the irradiation power at all times exceeds the accretion
power).

An additional mechanism to add material to the planet's atmosphere
involves capture of asteroids and planetesimals.
\citet{dong_et_al2010} point out that orbiting bodies that are
sufficiently small can be dynamically affected by passing through
stellar wind material, and can be dragged inward, changing their
period ratios with respect to the Jupiter-mass object.  If a
significant number of these are accreted, this process could affect
the bulk or atmospheric metallicity of the giant planet, and its
element ratios.  Note that changes in orbital eccentricity, of the
sort described by \citet{veras_et_al2011}, could also contribute to
small bodies having RGHJ-orbit-crossing trajectories.  Furthermore, in
the case of a binary star system, \citet{kratter+perets2012} show that
orbital instabilities can cause a planet to migrate significantly,
even to ``hop'' from one star to the other, which could again lead to
significant accretion of small rocky or icy material.

An evolved star that has gone through third dredge-up can have
significantly altered atmospheric abundances
\citep{kippenhahn+weigert1990}, and a C/O ratio greater than unity
even if on the main sequence its C/O ratio had been near solar
($\sim$0.5;
\citealt{asplund_et_al2009}).\footnote{\citet{lebertre_et_al2001} show
  that $\sim$50\% of mass loss from AGB stars is carbon-enriched.}  A
planet accreting such a star's wind, therefore, could appear to be a
carbon-enriched planet.\footnote{The planet's upper atmosphere might
  continue to remain carbon-enriched for as much as tens of millions
  of years after the end of the AGB phase (perhaps much less,
  depending on the strength of atmospheric mixing), after which the
  atmosphere will become well mixed with the deep interior and will no
  longer retain the memory of stellar wind pollution.}
\citet{madhusudhan_et_al2011b} found that multi-band infrared
photometry of WASP-12b \citep{hebb_et_al2009} suggests at high
confidence that this planet has a C/O ratio of 1 or greater, and
\citet{madhusudhan_et_al2011c} and \citet{oberg_et_al2011} explored
formation conditions that are necessary to form a carbon-rich hot
Jupiter around a main sequence star.  Here, we have shown that a
massive planet around an evolved star could acquire a carbon-enriched
atmosphere, even if the bulk interior of the planet has different
chemical abundances.  We note that a similar process might act to
change the composition of low-mass stellar companions (e.g., M stars
or white dwarfs) of AGB stars, as well \citep{livio+warner1984,
  farihi_et_al2010}.

\section{Chemical and Spectral Signatures}
\label{sec:chemspec}
The changes in the incident irradiation experienced by a Jupiter
around an evolving star cause distinct changes in chemical and thermal
properties in the planetary atmosphere that can be observed in
spectra. Here, we investigate the changes in chemistry and emergent
spectra of Jupiter as the Sun approaches the RGB phase. We model the
disk-integrated emergent spectra of Jupiter at various effective
temperatures using the radiative transfer code of
\citet{madhusudhan+seager2009}. The code solves line-by-line radiative
transfer in a 1-D plane-parallel atmosphere under the assumption of
hydrostatic equilibrium, local thermodynamic equilibrium (LTE), and
chemical equilibrium. For each planetary effective temperature we
consider, we adopt a 1-D averaged thermal profile of the planetary
atmosphere based on the analytic model of \citet{guillot2010}. We
choose model parameters for the thermal profile such that its thermal
gradient matches those of thermal profiles retrieved for exoplanets
with similar irradiation levels (e.g. GJ~436b;
\citealt{madhusudhan+seager2011}). We consider all the major sources
of opacity prevalent in H$_2$-dominated atmospheres, namely, line
absorption due to H$_2$O, CO, CH$_4$, CO$_2$, and NH$_3$, and
continuum opacity due to H$_2$-H$_2$ collision-induced absorption. For
a given thermal profile, we compute the mixing ratio profiles of the
molecular species under the assumption of chemical equilibrium
\citep{burrows+sharp1999}. We also compute the equilibrium
concentrations of the various species over a wide range of planetary
age, and hence thermal conditions, using the equilibrium chemistry
code developed in \citet{seager_et_al2005b}, and updated/used in
several following works (e.g. \citealt{miller-ricci_et_al2009a,
  madhusudhan+seager2011}).

\begin{figure}[t!]
\plotoneh
{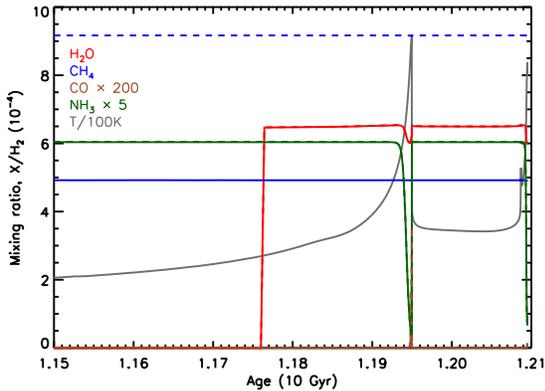}
\caption{Post-main-sequence evolution of Jupiter's equilibrium
  temperature and atmospheric chemical composition.  The temperature
  evolution is shown in gray (the gray curve is the planet's
  equilibrium temperature divided by 100~K, assuming that the star
  follows our adopted solar evolution model), and water, methane,
  carbon monoxide, and amonia mixing ratios are shown in red, blue,
  brown, and green, respectively.  Solid/dashed curves indicate
  abundances assuming carbon-to-oxygen ratio of 0.5/1.  These
  alternate C/O ratios illustrate possible changes in chemistry as a
  result of post-AGB atmospheric pollution or simply what might be
  found for planets with different primordial C/O ratios.  The main
  effect of doubling the C/O ratio is simply to double the CH$_4$
  abundance.  The evolution of the atmospheric water abundance is
  dramatic (and nearly independent of C/O ratio), from undetectable
  levels (as at present) to perhaps several times 10$^{-4}$ (depending
  on the as-yet unknown oxygen content of Jupiter) after the
  equilibrium temperature exceeds the condensation point of water.
  The dominant carbon-bearing species (CH$_4$ and CO) exhibit
  negligible temporal evolution in chemical equilibrium.}
\label{fig:chemistry}
\end{figure}

The changes in the chemical composition of the jovian atmosphere with
stellar age are shown in Fig.~\ref{fig:chemistry}. We discuss the
evolution of the molecular mixing ratios of the dominant species
(H$_2$O, CO, CH$_4$, and NH$_3$) in chemical equilibrium at a nominal
pressure of 1~bar. The changes in chemistry are driven largely by the
changing temperature of the atmosphere.

An important consequence is the enhanced observability of H$_2$O after
a stellar age ($t$) of $\sim$11.76~Gyr (i.e., after the equilibrium
temperature exceeds the sublimation point of water ice). At present,
the H$_2$O abundance in the jovian atmosphere is unknown. The low
temperatures ($T_{\rm eff} \sim 125$~K) cause H$_2$O to condense out
to the deep layers ($P \gtrsim 10$~bar) of the atmosphere, making
H$_2$O inaccessible to spectroscopic observations
\citep{atreya2004}. As the temperature increases beyond the
condensation temperature of H$_2$O, at $\sim$11.76~Gyr, gaseous H$_2$O
becomes abundant in the atmosphere and hence would be observable in
spectra from near-Earth space telescopes. H$_2$O then continues to be
the dominant oxygen-bearing molecule for the rest of the stellar
lifetime.\footnote{For temperatures of $T \sim 200$---900~K that are
  relevant in this context, H$_2$O is the dominant oxygen-bearing
  molecule and CH$_4$ is the dominant carbon-bearing molecule in
  H$_2$-dominated atmospheres, irrespective of the C/O ratio
  (e.g. \citealt{madhusudhan_et_al2011c}).}  A slight decrement in the
H$_2$O abundance is observed briefly as the star ascends the RGB ($t
\sim 11.95$~Gyr), when the planetary temperature exceeds $\sim$600~K;
at such temperatures, part of the oxygen ($\lesssim$10\%) is bound in
silicates.

As for carbon-bearing species, most of the carbon in the jovian
atmosphere continues to be in the form of CH$_4$, as is presently
known \citep{atreya2004, karkoschka+tomasko2010}. In chemical
equilibrium, CO is predicted to be negligible for most of the stellar
lifetime, and reaches a few ppm only at the peak temperatures
($\sim$900~K) when the star enters the RGB phase. However, our CO
estimate is only a lower limit, since it is known that non-equilibrium
chemistry in the form of vertical eddy diffusion can cause
enhancements in the CO abundance by over two orders of magnitude,
depending on the strength of large-scale vertical mixing
\citep{hubeny+burrows2007, madhusudhan+seager2011,
  visscher+moses2011}. In late stages of the evolution, substantial
accretion of stellar carbon-rich material onto the planetary
atmosphere is expected to enhance the C/O ratio of the planet. If
present-day Jupiter is oxygen-rich, with C/O = 0.5, such as the Sun, a
jovian atmosphere around an AGB star that has gone through third
dredge-up (see \S\ref{sec:acc}) can potentially have a C/O $\geq$1,
making it a Carbon-rich Planet (CRP;
\citealt{madhusudhan_et_al2011c}). As shown in
Fig.~\ref{fig:chemistry}, such an increase in the C/O ratio, due to
increased carbon, causes an increased CH$_4$ composition, while the
consequences on the remaining species remain minimal at the
temperatures under consideration.\footnote{The chemical consequences
  of high C/O ratios are more readily apparent, e.g. via a
  substantially decreased H$_2$O abundance, at $T \gtrsim 1200$~K
  \citep{madhusudhan_et_al2011c}.}

\begin{figure}[t!]
\plotoneh
{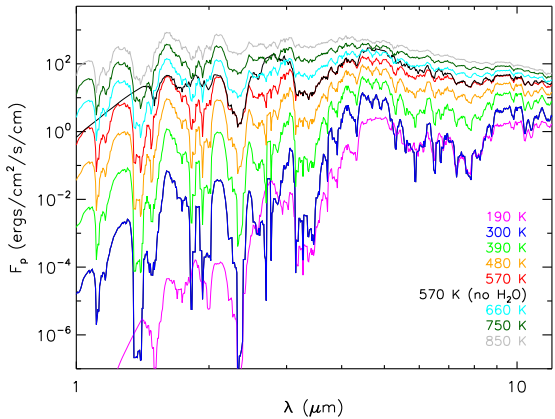}
\caption{Example spectra of Jupiter as a function of equilibrium
  temperature, as seen from a distance of 5~AU.  As Jupiter heats up,
  it becomes brighter by orders of magnitude in the near infrared part
  of the spectrum.  Methane features dominate the spectra at very low
  temperatures, and water features become prominent as the equilibrium
  temperature exceeds the sublimation point of water ice.  The red and
  the black curves are for the same equilibrium temperature (570~K),
  with and without H$_2$O, respectively.  The difference between them
  clearly indicates the spectral influence of atmospheric H$_2$O in
  the near infrared.}
\label{fig:spectra}
\end{figure}

The changes in chemistry and thermal irradiation cause concomitant
signatures in the emergent spectra of the jovian atmosphere at various
stages of the Sun's evolution. Figure~\ref{fig:spectra} shows emergent
spectra at 8 different effective temperatures, corresponding to
different solar ages. At very low temperatures ($T_{\rm eff} \lesssim
300$~K), absorption features of methane dominate the spectrum, and
water vapor is non-existent. As the jovian atmosphere continues to be
heated with increasing insolation, the blackbody continuum is enhanced
in the emergent spectrum. At the same time, features of water vapor
become conspicuous as atmospheric temperatures exceed $\sim$300~K. For
the rest of the solar evolution, features of H$_2$O and CH$_4$
dominate the spectrum. Consequently, in the distant future ($t \gtrsim
11.8$~Gyr), space telescopes will be able to derive the abundance of
water vapor in the jovian atmosphere from infrared
spectra.\footnote{At present, water vapor is under-abundant in the
  jovian atmosphere and hence not observable in infrared spectra. The
  Juno mission \citep{bolton2010} currently en route to Jupiter is
  expected to measure the water abundance in Jupiter using microwave
  sounding.}$^,$\footnote{At this distant-future epoch, of course, the
  Earth will be uninhabitable by life as we know it.} While the H$_2$O
and CH$_4$ abundances so derived from spectra would provide estimates
of the C/H and O/H elemental abundance ratios, a distinctive
determination of whether the atmospheric C/O ratio exceeds unity (and
hence is carbon-rich; \citealt{madhusudhan_et_al2011c}) would require
that the C/H and O/H ratios be estimated to precisions better than a
factor of $\sim$2.  The curves in Fig.~\ref{fig:spectra} represent
spectra as seen from a distance of 5~AU, and may be scaled to
represent fluxes from a hypothetical post-main-sequence Jupiter analog
around a star at any distance $d$ from us by multiplying the values by
$5.9 \times 10^{-10} (d/{1\rm~pc})^{-2}$.

\section{Atmospheric Motions}
\label{sec:motions}
The Rhines length scale is a measure of the typical size of zonal
bands of winds in a planetary atmosphere \citep{showman_et_al2010,
  vasavada+showman2005}.  This length scale, which describes the
boundary between turbulent (small) scales and the scales at which
coherent Rossby waves dominate, depends on flow speed $U$ as follows:
$L_{\rm Rh}[\beta] \sim \pi (U/\beta)^{1/2}$, where $\beta$ is the
northward derivative of the Coriolois parameter ($f\equiv 2\Omega
\sin[{\rm latitude}]$; and $\Omega$ is the planetary angular rotation
rate).  This length may be rewritten as an angular scale by dividing
by the planet's radius, yielding an equatorial Rhines angle that may
be expressed as
\begin{equation}
\theta_{\rm Rh} \sim 18\degr \left( \frac{U}{100 \rm~m/s} \times \frac{P_{\rm rot}}{1 \rm~day} \right)^{1/2} \left( \frac{R_p}{R_J} \right)^{-1/2} \, .
\label{eq:Rhines}
\end{equation}
Since post-main-sequence stellar evolution does nothing to increase
the strength of the stellar tide raised on a planet, RGHJs at
separations of $\sim$1~AU or more are not tidally locked to their
stars, and rotate at the same angular frequency they had during the
main-sequence portions of their stars' lives.  Their day lengths might
be $\sim$10~hrs if extrasolar Jupiters are similar to Jupiter and
Saturn, or perhaps even significantly less if they spin as fast as the
$\sim$2$-$3-hr periods that have been inferred for some brown dwarfs
\citep{clarke_et_al2002, artigau_et_al2009}.  We note that, lacking
permanent day and night, RGHJs probably have more uniform atmospheric
temperature distributions than their tidally locked bretheren around
main-sequence stars, and their heat redistribution parameter $P_n$
(describing the portion of day-side incident flux that is re-radiated
from the night side, per \citealt{burrows_et_al2006}) might be very
nearly 0.5.

It is difficult to know exactly how fast typical wind speeds on RGHJs
are likely to be, but it seems reasonable to assume they would be
between the speeds of winds on Jupiter ($\sim$40~m~s$^{-1}$) and those
on main-sequence hot Jupiters.  \citet{showman_et_al2010} and
\citet{menou2012} suggest two approximate scaling relations for how
hot Jupiter zonal wind speeds depend on day-night temperature
difference $\Delta T$.  If $v_{\rm rot,eq} \equiv \Omega_p R_p$ is the
equatorial rotation speed, then these two scaling relations can be
written as
\begin{equation}
U_1 \sim \left(\frac{k_B}{\mu m_p} N_{\rm SH} \Delta T\right)^{1/2} \quad \mbox{and} \quad U_2 \sim \frac{U_1^2}{v_{\rm rot,eq}} \, ,
\label{eq:v1}
\end{equation}
where $\mu$ is mean molecular weight in units of the proton mass and
$N_{\rm SH}$ is the number of scale heights over which there is a
vertical wind shear of $U_1$ or $U_2$, respectively.\footnote{Both
  expressions in Eq.~(\ref{eq:v1}) are found in
  \citet{showman_et_al2010} and the latter (for ``$U_2$'') is found in
  \citet{menou2012}.}
For a hot Jupiter, $v_{\rm rot,eq}$, $U_1$, and $U_2$ all are of order
one or a few km~s$^{-1}$.  For a RGHJ, $v_{\rm rot,eq}$ is an order of
magnitude greater, so, over $N_{\rm SH}=3$ scale heights, $U_1 \sim
1000 {\rm~m~s^{-1}}(\Delta T/100 {\rm~K})^{1/2}$, while $U_2\sim 100
{\rm~m~s^{-1}} (\Delta T/ 100 {\rm~K})$.

In Table~\ref{ta:scale}, we provide a comparison of the angular Rhines
scale for Jupiter, an ordinary hot Jupiter, and a RGHJ.  Dynamical
models suggest that wind speeds in hot Jupiter atmospheres are roughly
a kilometer per second (\citealt{showman_et_al2008b,
  rauscher+menou2009, heng_et_al2011}, and many others).  These
extremely fast winds are largely driven by the strong day-night
temperature contrast imposed by tidally locked rotation, and tend to
be organized into large-scale jets (typical angular Rhines scale of
$\sim$100$\degr$).  A RGHJ, lacking this strong temperature contrast,
might have significantly slower winds.  In Table~\ref{ta:scale}, we
have assumed $100\rm~m~s^{-1}$ as an illustrative value.  Since the
angular Rhines scale depends on wind speed to the $1/2$ power, errors
of order unity in wind speed do not change the qualitative conclusion
that the combination of faster rotation and slower winds would lead
the general character of atmospheric dynamics on a RGHJ to be more
similar to that of Jupiter, with many narrow banded jets.

\begin{table}[t]
\small
\begin{center}
\caption{Scale of Atmospheric Features} \label{ta:scale}
\begin{tabular}{l|cccc}
\hline
\hline
                         &             &              &        &                  \\[-0.2cm]
\multirow{2}{*} {Object} & $U$         & $P_{\rm rot}$ & $R_p$   & $\theta_{\rm Rh}$ \\
                         & (m~s$^{-1}$) & (days)       & ($R_J$) &                 \\
\hline
\rule {-3pt} {10pt}
Jupiter               & $\sim$40       & 0.4          & 1    & $\sim$7$^\circ$   \\[0.2cm]
Hot Jupiter           & $\sim$10$^3$   &  4           & 1.3  & $\sim$100$^\circ$ \\[0.2cm]
RGHJ                  & $\sim$10$^2$?  & 0.4          & 1    & $\sim$10$^\circ$  \\[0.2cm]
\end{tabular}
\label{ta:Kzz_a}
\tablecomments{This table presents typical values of wind-speed ($U$),
  rotation period ($P_{\rm rot}$), planetary radius ($R_p$), and
  angular Rhines scale ($\theta_{\rm Rh}$) for Jupiter, for a hot
  Jupiter around a main-sequence star, and for a red-giant hot Jupiter
  (RGHJ) that is too far from its star to be tidally locked.  The
  small angular Rhines scale for the RGHJ suggests that it would have
  narrow-banded jets, similar to Jupiter and different from what
  dynamical models suggest for ordinary hot Jupiters.}
\end{center}
\end{table}

\section{Conclusions}
\label{sec:conc}
It has long been appreciated that as a star expands and grows more
luminous after its main-sequence lifetime, the orbital radii
corresponding to a given equilibrium temperature move farther out from
the star.  \citet{lopez_et_al2005} and
\citet{schroder+connonsmith2008}, for instance, considered the outward
expansion of the habitable zone after a star leaves the main sequence,
although whether life would be likely to have enough time to develop
from abiotic conditions in the limited time available remains an open
question \citep{spiegel+turner2012}.  A large number of known
exoplanets are far enough from their stars that they are not currently
strongly irradiated but they will be in the future.  To investigate
the properties that such planetary companions will have in the future,
when their stars are much more luminous, we computed a suite of
post-main-sequence stellar models to see how far out an object could
be and still plausibly be called a hot Jupiter around a sufficiently
evolved star.  We found that, as far out as $\sim$35~AU from a
3-$M_\sun$ star, a companion might transiently be highly enough
irradiated to merit the moniker ``hot Jupiter.''  We argued that
massive planets might accrete an amount of stellar wind that is large
compared with their own atmospheres, resulting in significant
atmospheric pollution if the stellar wind's composition is different
from the planet's, and perhaps creating the atmospheric signature of a
CRP (even if the planet's bulk composition is not carbon-rich).  We
showed that an angular-Rhines-scale argument suggests that that RGHJs
might be expected to have narrow-banded atmospheric jets, in contrast
to main-sequence hot Jupiters.

Detection of RGHJs might prove to be observationally challenging.  The
probability that a planet transits its star increases dramatically as
the star's radius increases by a factor of 100 or more, but
recognizing transits (or occultations) of RGHJs could prove nearly
impossible, since the transit depth becomes tiny ($\sim$10$^{-6}$) and
the duration could last weeks to months \citep{assef_et_al2009}.
High-contrast imaging is a potential avenue for detecting photons from
RGHJs, but the solar neighborhood is not rich in likely suitable
targets and the required angular resolution to resolve companions to
distant giant stars might be daunting, given present technology.  It
might actually be easier to find evidence of former RGHJs than of
current RGHJs, perhaps by surveying white dwarfs for companions at
several AU \citep{gould+kilic2008}.  Purely because of the timescale
of photon diffusion, energetic arguments suggest that a companion
should remain hot, after an AGB star fades into oblivion, for roughly
as long as it had been highly irradiated (a few hundred million
years).  Such a reheated companion, perhaps identified by its infrared
excess relative to the white dwarf's spectral energy distribution,
might be mistaken for a more massive companion.  However, this
potential source of confusion is quantitatively significant only for
very young white dwarfs, since the thermal relaxation timescale of a
reheated planet is also of order a few hundred million years, short in
comparison to a Hubble time.

The searches for exoplanets in the last two decades have uncovered
many surprises.  Here, we investigated the properties of a category of
giant planets that might be found soon and that have not been studied
in detail previously.  Hot Jupiters around post-main-sequence stars
must exist in the Galaxy and represent a future evolutionary stage of
many of the known jovian-mass planets, including our own Jupiter.

\vspace{0.5in}


{\sc Acknowledgments}

We thank many people for useful discussions, in particular Rodrigo
Fernandez, JJ Hermes, Ruobing Dong, Scott Tremaine, Ed Turner, Jason
Nordhaus, Adam Burrows, Jeremy Goodman, Doug Lin, Subo Dong, Jay
Farihi, Jonathan Mitchell, Adam Showman, Kristen Menou, Scott Gaudi,
and John Johnson.  We also thank an anonymous referee for helpful
comments that materially improved the manuscript.  DSS gratefully
acknowledges support from NSF grant AST-0807444 and the Keck
Fellowship.  NM acknowledges support from the Yale Center for
Astronomy and Astrophysics through the YCAA postdoctoral Fellowship.


\bibliography{biblio.bib}

\clearpage

\end{document}